\documentstyle[preprint,aps,tighten]{revtex}

\input{psfig}

\begin{document}

\draft
\preprint{}

\title{Analysis on the diffractive production of $W$'s and dijets 
at the DESY HERA and Fermilab Tevatron colliders\footnote{To be published in 
Physical Review D.}}

\author{R. J. M. Covolan and M. S. Soares\\
{\small Instituto de F\'{\i}sica {\em Gleb Wataghin}} \\ {\small Universidade
Estadual de Campinas, Unicamp} \\ {\small 13083-970 \  Campinas \  SP
 \
Brazil \bigskip}}
\date{\today}
\maketitle

\begin{abstract}
Hadronic processes in which hard diffractive production takes place
have been observed and analyzed in collider experiments for several
years. The ex\-pe\-rimental rates of diffractive $W$'s and dijets
measured at the Tevatron and the cross sections of diffractively
produced dijets recently obtained at the HERA experiment are the
object of this analysis.  We use the Pomeron structure function
obtained from the HERA data by two different approaches to calculate
the rates and cross sections for these processes. The comparison of
theoretical predictions with the measured values reveals some
discrepancies that make evident conceptual difficulties with such
approaches.  A new version of the Ingelman-Schlein model is proposed
as an attempt to overcome such difficulties and make theory and data
compatible.

\vspace{1cm}

PACS: {12.40.Nn, 13.60.Hb, 13.60.-r, 13.85.Qk, 13.87.-a}

\end{abstract}

\newpage

\section{Introduction}

The phenomenological analysis of hard diffractive processes has
become one of the most interesting theoretical laboratories to
investigate the nature and structure of the Pomeron. The concept of
Pomeron structure function was introduced by Ingelman and Schlein
\cite{ingelman} as an ansatz to investigate the eventual production
of high-$p_T$ jets in diffractive hadron interactions. Such a
theoretical speculation has become reality when the UA8 Collaboration
obtained the first measurements of diffractively produced dijets
\cite{ua8}.  However, further quantitative analyses \cite{schlein}
carried out at the CERN ${\bar p} p$ Collider have shown that  the
predicted rates obtained by the Ingelman-Schlein (IS) model were too
much high in comparison with the measured values.

The same problem appeared when similar  measurements were performed
at the Fermilab Tevatron Collider. The ratio of diffractive to
non-diffractive dijets predicted by the IS model \cite{bruni} for the
Tevatron energy resulted to be more than one order of magnitude above
the actual measurements.

Despite the fact that the IS predictions about the existence of
diffractive jets turned out to be correct, one might think that this
model is acceptable only in a ``qualitative'' sense since the
predicted rates were completely off the values actually measured. Let
us examine this issue in a more detailed way.

The calculational scheme proposed by such a model makes use of the
so-called factorization hypothesis, that is in a reaction like $p +
p \rightarrow p + jets + X$, the whole process is supposed to occur
as a sequence of two {\it independent} steps: first a proton emits a
Pomeron, then partons of the  Pomeron interact with partons of the
other proton producing (for instance) jets. This picture seems to be
a sort of straightforward extension of the parton model to 
diffractive processes. In fact, it is quite appealing as such, but it
is not obvious that the factorization property should apply to this
type of hadron interactions (see \cite{alvero} and references
therein).

A similar kind of problem affects also another class of processes.
Experimental results apparently in favor of factorization
have come up from the DESY $e p$ Collider, the so-called HERA
experiment. Events of deep inelastic scattering (DIS) tagged with
rapidity gaps, observed and analyzed by the H1 and ZEUS
collaborations \cite{h1,zeus}, have shown the same characteristic
pattern exhibited by hadron diffractive dissociation processes in the
kinematical region where Pomeron exchange is dominant.  These
observations strongly suggested that, in fact, the internal structure
of the Pomeron was being probed.  Moreover, this pattern, which
resembles the peak observed in soft diffractive dissociation, seemed
to be independent on kinematical variables other than $x_{\tt I\!
P}$ (the fraction of the proton momentum carried by the Pomeron),
allowing that estimations of the Pomeron intercept were done
\cite{h1,zeus}.

These were the first results of diffractive structure function.
More recently, further and more precise measurements performed over
an extended kinematical region have shown evidences of factorization
breaking in the $x_{\tt I\! P}$ variable \cite{h1_novo}. Such an
effect, however, can be understood by an adequate computation of
secondary-reggeon contributions \cite{h1_novo}, leaving open the
possibility that factorization applies when Pomeron exchange is the
dominant mechanism.

Nevertheless, this conclusion has to be taken with some caution
because other forms of factorization breaking, which are not evident
from the $x_{\tt I\! P}$ dependence, can occur. For the moment, let
us assume (as the IS model does) that the factorization hypothesis
is an acceptable ansatz. This assumption leads us to more definite
theoretical questions which were present in the IS model from the
very beginning.

From a quantitative point-of-view, factorization appears in the IS
model as the product of two quantities representing the two-step
process mentioned above: (1) the so-called Pomeron {\it flux factor},
which is supposed to give the probability of a Pomeron being emitted
by a proton (or antiproton), and (2) the elementary cross section
resulting from the interaction among partons belonging to the
Pomeron and to the other proton \cite{ingelman}. In order to
calculate this elementary cross section, the knowlegde of the
Pomeron structure function is, of course, an indispensable requisite.
By the time the IS model was proposed there was no experimental
information about the Pomeron structure and, thus, only estimations
based on ``educated guesses'' could be done \cite{ingelman,bruni}. 
Today, the HERA data of diffractive DIS are available and one can try
to extract from them the Pomeron structure function. However, this is
not a straightforward procedure because the results are dependent on
the model one chooses for the Pomeron flux factor.

In a recent paper \cite{covolan}, we have presented a study on the
Pomeron structure function in which two different forms of flux
factor were employed, one derived from the standard Regge theory
\cite{donna} and the other obtained from the so-called {\it
renormalization} procedure \cite{dino}. The former assumes
factorization whereas the latter implies in a sort of factorization
breaking.  We have shown that the quark/gluon content of the Pomeron
changes significantly whether one chooses the standard or the
renormalized flux factor. In the present paper, we apply these
results to estimate the rates of diffractively produced $W$'s and
jets and compare such estimations with the available data.

As we shall see, no one of these models, standard or renormalized, is
able to offer a completely satisfactory description of the data,
although the latter is partially successfull. In order to overcome
these difficulties, we introduce a new version of the IS model, which
is quite intuitive and presents promising results.

Although the comparison to data is central to our analysis,
we emphasize that no attempt to {\it fit} the theoretical outcomes to
the experimental rates or cross sections has been done\footnote{We
refer the reader interested in appreciating an analysis that tries
to conciliate the IS model with data by fitting, to the paper of
Ref.~\cite{alvero}.}. In fact, such an attempt could conceal the
problems that we intend to make evident.

The paper is organized as follows. In Sec.~II, we summarize our
procedure to determine the Pomeron structure function from HERA data
and present the parametrizations obtained in \cite{covolan} that are
used in this paper. In Sec.~III, we present the formalism used to
calculate the cross sections for diffractive production of $W$'s and
jets. In Sec.~IV, the results obtained with the standard and
renormalized flux factors are shown in comparison with the
experimental data. In Sec.~V, we present a discussion about the IS
model and a new approach is suggested. Our concluding remarks are in
Sec.~VI.

%
%

\section{Pomeron structure function from HERA data}

The cross section for {\it diffractive} DIS processes,
\begin{equation}
e (k) + p (P) \rightarrow e' (k') + p' (P') + X (M_X^2),
\label{proc}
\end{equation}
is given by the expression 
\begin{eqnarray}
\frac{d^{4}\sigma^{D}}{dx\ dQ^{2}\ dx_{ {\tt I\! P}}\ dt}=\frac{4\pi
\alpha^{2}}{x\ Q^{4}}\ \left\{1-y+\frac{y^{2}}{2[1+R^{D}(x,Q^{2},x_{
{\tt I\! P}},t)]}
\right\}\ F^{D(4)}_{2}(x,Q^{2},x_{\tt I\! P},t),
\end{eqnarray}
where $x, Q^2$ and $y$ are the usual DIS variables. Besides these,
two other 
variables are used to specify process (\ref{proc}),
\begin{equation}
x_{\tt I\! P} = \frac{M_X^2 + Q^2 -t}{W^2 + Q^2 - m^2_p} \cong
\frac{M_X^2 + Q^2}{W^2 + Q^2},
\label{xpom}
\end{equation}
and
\begin{equation}
\beta = \frac{Q^2}{M_X^2 + Q^2 -t} \cong \frac{Q^2}{M_X^2 + Q^2},
\label{beta}
\end{equation}
where $x_{\tt I\! P}$ is the momentum fraction carried by the Pomeron
emitted by the scattered proton and $\beta$ is the fraction of the 
Pomeron momentum carried by  the struck quark. In Eqs. (\ref{xpom})
and (\ref{beta}), $W$ is the  energy in the $\gamma^* p$
center-of-mass system,  $t=(P-P')^2$ is the four-momentum  transfer
at the proton vertex and $M_X$ is the invariant mass of the hadronic 
system $X$. From these equations, one obtains the relationship
\begin{equation}
x_{\tt I\! P} = \frac{x}{\beta}.
\label{xp}
\end{equation} 

The HERA data \cite{h1,zeus} used to perform our analysis \cite{covolan}
were obtained under the assumption that $R^{D}(x,Q^{2},x_{ 
{\tt I\! P}},t) = 0$ and, since $t$ was not measured, the cross
section must be considered as integrated over this variable, that is
\begin{equation}
\frac{d^{3}\sigma^{D}}{d{\beta}\ dQ^{2}\ dx_{ 
{\tt I\! P}}}=\frac{2\pi
\alpha^{2}}{\beta\ Q^{4}}\ [1+(1-y)^2]\
F^{D(3)}_{2}(\beta,Q^{2},x_{\tt I\! P}),
\end{equation}
so that the experimental data are expressed in terms of the {\it
diffractive} structure function $F^{D(3)}_{2}(\beta,Q^{2},
x_{\tt I\! P})$.

These data have shown for the first time a very clear diffractive
pattern, that is the characteristic $x_{\tt I\! P}$-dependence
observed in diffractive dissociation of hadrons. This feature was
observed irrespective of the ($\beta, Q^2$) values considered and
this was very suggestive that a factorized expression like
\begin{eqnarray} 
F_{2}^{D(3)}(\beta,Q^{2},x_{\tt I\! P})=g(x_{\tt I\! P})\ F_{2}^
{ {\tt I\! P}}(\beta, Q^{2}) 
\label{Fdiff} 
\end{eqnarray} 
could apply\footnote{As mentioned in the Introduction, new data
obtained by the H1 Collaboration \cite{h1_novo} in  more extensive
kinematical regions of both $\beta$ and $Q^2$ have shown very clear
factorization breaking which, however, can be understood as being an
effect of secondary reggeon contributions other than the Pomeron. 
Our analysis \cite{covolan} refers to data obtained in a particular
kinematical region where Pomeron exchanges are supposed to be
dominant.}. Based on this factorization hypothesis and on the IS
model \cite{ingelman}, it is usual to interpret $g(x_{\tt I\! P})$ as
the integrated-over-$t$  Pomeron flux factor and 
$F_{2}^{\tt I\! P}(\beta, Q^{2})$ as the Pomeron structure function.

Besides our analysis, there are others in the literature that are
based on a similar procedure (see for instance
\cite{alvero,gehrmann,golec}). Our main concern, however, was
confronting two different approaches: one in which the {\it standard}
flux factor is employed and the other which corresponds to the {\it
renormalized} flux factor (for brevity, we will refer to these
quantities hereafter as STD and REN flux factors, respectively). For
the former, it was assumed the Donnachie-Landshoff expression
\cite{donna}, 
\begin{eqnarray}   f_{STD}(x_{\tt I\! P},t)=\frac{9\beta_0^2}{4\pi^{2}}
\ [F_{1}(t)]^{2}\ 
x_{\tt I\! P}^{1-2\alpha(t)},
\label{dlflux} 
\end{eqnarray}
whereas the latter is determined from the procedure prescribed in
\cite{dino}, that is 
\begin{equation}
f_{REN}(x_{\tt I\! P},t)=\frac{f_{STD}(x_{\tt I\! P},t)}
{N(x_{{\tt I\! P}_{min}})}
\label{diflux}
\end{equation}
where 
\begin{equation}
N(x_{{\tt I\! P}_{min}})=\int_{x_{{\tt I\! P}_{min}}}^{x_{
{\tt I\! P}_{max}}}dx_{\tt I\! P}
\int^{0}_{t=-\infty}f_{STD}(x_{\tt I\! P},t)\ dt.
\label{norm}
\end{equation}
By introducing Eq.~(\ref{dlflux}) into Eq.~(\ref{norm}) and assuming
an exponential approximation for the form factor, $F_{1}^2(t) \simeq
e^{b_{0}(t)}$, one obtains
\begin{equation}
N(x_{{\tt I\! P}_{min}}) = K\ \frac{e^{-\gamma}}{2\alpha'}\
[E_i(\gamma - 2\epsilon\ \ln x_{{\tt I\! P}_{min}}) - E_i(\gamma -
2\epsilon\ \ln x_{{\tt I\! P}_{max}})]
\label{renorm}
\end{equation}
where $E_i (x)$ is the exponential integral, $K =
{9\beta_{0}^{2}}/{4\pi^{2}}$ and $\gamma = b_0 \epsilon/\alpha'$.

An important point to be noted here is that, in the calculation of
the soft diffractive dissociation cross section, the minimum value
of $x_{\tt I\! P}$ is $x_{{\tt I\! P}_{min}} = (m_p + m_{\pi})^2/s$,
but it is assumed  that when one applies 
Eqs.~(\ref{diflux})-(\ref{renorm}) to
the diffractive DIS analysis this quantity should become (see
\cite{dino})
\begin{equation}
x_{{\tt I\! P}_{min}} = \frac{Q^2}{\beta\ s}.
\label{xmin}
\end{equation}
This distinction is pretty important and will be a matter of
discussion in Section IV.

With these flux factors (integrated over $t$) introduced into
Eq.~(\ref{Fdiff}), the Pomeron structure function was obtained from
$F_{2}^{D(3)}(\beta,Q^{2},x_{\tt I\! P})$ data under the assumption
that 
\begin{equation}
F^{\tt I\! P}_{2}(\beta,Q^{2}) = \sum_{i=u,d,s} e^2_i\ \beta\
[q_i(\beta, Q^2) + {\bar q_i} (\beta, Q^2)] = \frac{2}{9}\ \beta\
\Sigma (\beta, Q^2),
\label{pomsf}
\end{equation} 
with $\Sigma (\beta, Q^2) = \sum_{i=u,d,s} \ [q_i(\beta, Q^2) + 
{\bar q_i}(\beta, Q^2)]$ representing a quark singlet that would
evolve with $Q^2$ according to the DGLAP equations \cite{dglap}. The
starting scale for the evolution was assumed to be $Q^{2}_{0}=4$
GeV$^2$ and no sum rule was imposed on the parametrizations to
perform the fitting (however, in one of the cases presented below, we
use the sum rule to determine a parameter of normalization).

Details about this procedure and about the results can be found in
\cite{covolan}, but roughly speaking this analysis has put in
evidence three major points:

\begin{itemize}

\item The quark/gluon content of Pomeron as obtained from HERA data
via IS model depends strongly on which kind of flux factor is
assumed. That means that the issue about (re)normalization is
crucial.

\item STD flux factor favors a predominance of gluons at the initial
scale of $Q^2$ evolution while basically the contrary happens with
the renormalized scheme.

\item Several different trials were carried out during the fitting
procedure. In almost all of them, the initial quark and gluon
distributions preferred a hard or super-hard shape. Thus, this
analysis has practically ruled out the possibility for soft
distributions at initial scale in diffractive DIS.

\end{itemize}

For the present analysis, we have chosen two parametrizations for
each flux factor.  These parametrizations, the most representative of
our analysis,  are described below.

\noindent{\bf Fit~1}: These parametrizations were obtained in
\cite{covolan} with STD flux and correspond to a combination that we
call {\it hard-hard}, i.e. both quark and gluon distributions have a
hard shape at the initial scale of evolution:  
\begin{eqnarray} 
\nonumber
\beta\ \Sigma( \beta,Q^{2}_{0})&=&2.55\ \beta\ (1- \beta),\\ 
\beta\ g(\beta,Q^{2}_{0})&=&12.08\ \beta\ (1- \beta).  
\label{fit1}
\end{eqnarray}

\noindent{\bf Fit~2}: This case refers also to parametrizations
obtained with the STD flux; the initial distributions correspond to a
super-hard profile imposed to gluons by a delta function while quarks
were left free to change according to the data (we refer to this case
as {\it free-delta}):
\begin{eqnarray} 
\nonumber 
\beta\ \Sigma(\beta,Q^{2}_{0})&=&1.51\ \beta^{0.51}\ (1-
 \beta)^{0.84},\\ 
\beta\ g(\beta,Q^{2}_{0})&=&2.06\ \delta(1- \beta).  
\label{fit2} 
\end{eqnarray}

\noindent{\bf Fit~3}: These are parametrizations obtained with the
REN flux factor and a initial combination of the type {\it
hard-hard}:
\begin{eqnarray} 
\nonumber 
\beta\ \Sigma(\beta,Q^{2}_{0})&=&5.02\ \beta\ (1- \beta),\\ 
\beta\ g(\beta,Q^{2}_{0})&=&0.98\ \beta\ (1- \beta).  
\label{fit3}
\end{eqnarray}

\noindent{\bf Fit~4}: These are also parametrizations obtained with
the REN flux factor and a combination of the type {\it free-zero},
that is the quark distribution was left free while the initial gluon
distribution was imposed to be null:
\begin{eqnarray} 
\nonumber 
\beta\ \Sigma(\beta,Q^{2}_{0})&=&2.80\ \beta^{0.65}\ (1-
 \beta)^{0.58},\\ 
\beta\ g(\beta,Q^{2}_{0})&=&0. 
\label{fit4} 
\end{eqnarray}

In the case of Fit~3, it was very difficult to fix the normalization
parameter for the gluon distribution (see \cite{covolan}). In the
expression used here, Eq.~(\ref{fit3}), this parameter was
established by using the normalization obtained for quarks and
imposing sum rule. In fact, it was because  of such a difficulty
that Fit~4 was performed.

All these four combinations of flux factors and Pomeron structure
functions were applied in the calculation that follows.

%
%

\section{Cross Sections for Diffractive Production Processes}

As we have seen in the previous section, several possible
parametrizations for the Pomeron structure function are allowed by
the HERA data depending on what one assumes for the flux factor. Our
aim is applying these parametrizations to calculate the diffractive
production rates of different processes in order to  compare the 
results with the available data and analyze the implications. In this
section, we present the cross sections used to perform these
calculations.

The generic cross section of a process in which partons of two
hadrons, A and B, interact to produce jets (or $W$'s),
\begin{equation}
A + B \rightarrow \ Jets\ (W) + X, 
\end{equation}
is given by the standard parton model as
\begin{equation}
\label{gen}
d \sigma_{A B \rightarrow  Jets (W)} = \sum_{a,b,c,d} f_{a/A}(x_a,
 \mu^2)\ dx_a\ f_{b/B}(x_b, \mu^2)\ dx_b\
 \frac{d\hat{\sigma}_{ab\rightarrow  cd (W)}}{d\hat{t}}\ d\hat{t}.
\end{equation}

In order to adapt such a cross section to a hard diffractive
interaction, one assumes (in the spirit of the IS model) that one of
the hadrons, say hadron $A$, emits a Pomeron which is made up of
partons itself. The procedure we adopt in such a case is replacing
$x_a f_{a/A}(x_a, \mu^2)$ in Eq.~(\ref{gen}) by the convolution
between the distribution of partons in the Pomeron, $\beta 
f_{a/{\tt I\! P}}(\beta, \mu^2)$, and the ``emission rate"
of Pomerons by $A$, $f_{\tt I\! P}(x_{\tt I\! P},t)$. The first
quantity corresponds to distributions like those discussed in the
previous section, that take part in the definition of the Pomeron
structure function, while the second corresponds to the Pomeron flux
factor.  In such a case, we have 
\begin{eqnarray}
\label{convol}
x_a\ f_{a/A}(x_a, \mu^2)\ =\ \int dx_{\tt I\! P} \int d\beta \int dt\
 f_{\tt I\! P}(x_{\tt I\! P},t)\ \beta\ 
f_{a/{\tt I\! P}}(\beta, \mu^2)\ \delta(\beta-{x_a}/{x_{\tt I\! P}}).
\end{eqnarray}
Defining  $g (x_{\tt I\! P}) \equiv \int_{-\infty}^0 dt\ 
f_{\tt I\! P}(x_{\tt I\! P},t)$, one obtains 
\begin{equation}
\label{convoP}
x_a\ f_{a/A}(x_a, \mu^2)\ =\ \int \frac{dx_{\tt I\! P}}
{x_{\tt I\! P}}\ g (x_{\tt I\! P})\ {x_a}\ 
 f_{a/{\tt I\! P}}({x_a}/{x_{\tt I\! P}}, \mu^2).
\end{equation}
This is the characteristic procedure that is applied below to
calculate the cross section of diffractive processes.

%
%

\subsection{Diffractive Hadroproduction of $W^{\pm}$}

In this analysis, we consider $W^{\pm}$ diffractive production in the
 reaction
\begin{equation}
\label{reacao}
p + {\bar p} \rightarrow p + \ W (\rightarrow e\ \nu ) + \ X,
\end{equation}
which was experimentally studied at the Tevatron Collider by the CDF
Collaboration \cite{wcdf}. We assume that a Pomeron emitted by a
proton in the  positive $z$-direction interacts with a $\bar p$
producing $W^{\pm}$ that subsequently decay into $e^{\pm}\ \nu$. In
this configuration, the detected lepton ($e^+$ or $e^-$) would
appear boosted towards negative $\eta$ (rapidity) in coincidence
with a rapidity gap in the right hemisphere. 

The cross section for the inclusive lepton production by this process
is\footnote{See a more detailed discussion about this cross section
in the Appendix.}
\begin{eqnarray}
\frac{d\sigma}{d\eta_e}= \sum_{a,b} \int dx_{\tt I\! P}\ 
g(x_{\tt I\! P})\int dE_T \ f_{a/{\tt I\! P}}(x_a, \mu^2)\ 
f_{b/\bar{p}}(x_b, \mu^2)\ \left[\frac{ V_{ab}^2\ G_F^2}{6 s 
\Gamma_W M_W}\right]\  \frac{\hat{t}^2}{\sqrt{A^2-1}}
\label{dsw}
\end{eqnarray}
where
\begin{equation}
x_a  = \frac{M_W\ e^{\eta_e}}{(\sqrt{s}\ x_{\tt I\! P})}\ \left[A 
\pm \sqrt{(A^2-1)}\right],
\label{xaw}
\end{equation}
\begin{equation}
x_b = \frac{M_W\ e^{-\eta_e}}{\sqrt{s}}\ \left[A \mp
 \sqrt{(A^2-1)}\right],
\label{xbw}
\end{equation}
and
\begin{equation}
\hat{t}=-E_T\ M_W\ \left[A+\sqrt{(A^2-1)}\right]
\label{tw}
\end{equation}
with $ A={M_W}/{2 E_T}$. The upper signs in Eqs.~(\ref{xaw}) and
(\ref{xbw}) refer to $W^+$ production (that is, $e^+$ detection). The
corresponding cross section for $W^-$ is obtained by using the lower
signs and ${\hat t} \leftrightarrow {\hat u}$ (see the Appendix).

Since the experimental data of diffractive $W$ production presently 
available 
are not highly precise (in fact, CDF data \cite{wcdf} are the first 
measurements of this process), our calculations consider only leading 
order contributions.
In this sense, we note that studies of non-diffractive $W$ production 
performed 
with CDF cuts  indicate that next-to-leading (NLO) order 
corrections could be around 10\% of the production cross section (see 
 \cite{baer}).
  However, the experimental information is given in terms of a ratio
between diffractive to non-diffractive $W$ production, thus the effect of 
NLO corrections in the final results are expected to be even smaller than 
this percentage.

The contribution of other competitive processes like inclusive 
hadroproduction of $W + jet$ and $W + \gamma$ (and respective 
NLO corrections) are expected to be very small as well (about the former 
process, see for instance analyses by Giele {\it et al.} in Ref.\cite{giele}
and references therein,  
and the CDF account \cite{wcdf}, while with respect to the latter we refer 
the reader to the work of U. Baur {\it et al.} in Ref.~\cite{baur}).

%
%

\subsection{ Diffractive Hadroproduction of Dijets}

In order to calculate the process by which two hadrons, say $p$ and
$\bar p$, interact generating dijets,
\begin{equation}
p + {\bar p} \rightarrow \ j_1+j_2+X, 
\end{equation}
one starts from the expression
\begin{equation}
\label{modp}
d \sigma_{p {\bar p} \rightarrow  j_1 j_2} = \sum_{a,b,c,d}
 f_{a/p}(x_a, \mu^2)\ dx_a\ f_{b/{\bar p}}(x_b, \mu^2)\ dx_b\ 
 \frac{d\hat{\sigma}_{ab\rightarrow  cd}}{d\hat{t}}\ d\hat{t},
\end{equation}
in which it is assumed that partons $a$ and $b$ belong to hadrons $p$
and $\bar p$ in their initial states, and partons $c$ and $d$ will
give rise (in leading order) to the dijet pair in the final state.
The distributions $f_{a/p}(x_a, \mu^2)$ and $f_{b/{\bar p}}(x_b,
\mu^2)$ are the structure functions evolved to an adequate scale
$\mu^2$ and $d\hat{\sigma}_{ab\rightarrow  cd}/{d\hat{t}}$ stands
for the QCD matrix elements proper to this calculation.

With a suitable changing of variables, namely $dx_a\ dx_b\ d\hat{t}
\rightarrow 2E_T\ x_a\ x_b\ dE_T\ d\eta'\ d\eta$, one gets the
differential cross section in terms of the rapidity $\eta$ of one of
the jets,
\begin{eqnarray}
\label{modpar}
\frac{d\sigma}{d\eta}=\sum_{a,b,c,d}\int_{E_{T_{min}}}^{E_{T_{max}}}
 dE_T^2 \int_{\eta'_{min}}^{\eta'_{max}} d\eta' \ x_a\ f_{a/p}(x_a,
 \mu^2)\ x_b\ f_{b/{\bar p}}(x_b, \mu^2)\ 
 \frac{d\hat{\sigma}_{ab\rightarrow  cd}}{d\hat{t}},
\end{eqnarray}
where
\begin{equation} 
x_a =  \frac{E_T}{\sqrt{s}}(e^{-\eta}+e^{-\eta'}),
\ \ \ \ \ \ \ 
x_b = \frac{E_T}{\sqrt{s}}(e^{\eta}+e^{\eta'}), 
\label{xbj}
\end{equation}
with $E_T$ being the jets transversal energy. These expressions apply
to the non-diffractive case.

By using the convolution procedure described above, the cross section
 for diffractive hadroproduction of dijets becomes 
\begin{eqnarray}
\frac{d\sigma}{d\eta}=\sum_{a,b,c,d}\int_{E_{T_{min}}}^{E_{T_{max}}}
 dE_T^2 \int_{\eta'_{min}}^{\eta'_{max}} d\eta' 
\int_{x_{{\tt I\! P}_{min}}}^{x_{{\tt I\! P}_{max}}} dx_{\tt I\! P}
\ g  (x_{\tt I\! P})\ \beta f_{a/{\tt I\! P}}(\beta, \mu^2)
\ x_bf_{b/\bar{p}}(x_b, \mu^2)\ \frac{d\hat{\sigma}_{ab\rightarrow
  cd}}{d\hat{t}},
\label{dsigjato} 
\end{eqnarray}
where $\beta = {x_a}/x_{\tt I\! P}$ with $x_a$ and $x_b$ given by 
(\ref{xbj}).
The kinematical limits for the above expression are
\begin{equation}
\ln{\frac{E_T}{\sqrt{s}-E_T\ e^{-\eta}}} \leq \eta' \leq 
\ln{\frac{\sqrt{s}-E_T\ e^{-\eta}}{E_T}},
\label{limeta}
\end{equation}
and
\begin{equation}
E_{T_{max}}=\frac{\sqrt{s}}{e^{-\eta}+e^{\eta}},
\label{limet}
\end{equation}
with 
$E_{T_{min}}$, $x_{{\tt I\! P}_{min}}$, and $x_{{\tt I\! P}_{max}}$ 
established by experimental cuts.

%
%

\subsection{Diffractive Photoproduction of Dijets}

The process considered now is diffractive photoproduction of dijets,
obtained from the reaction 
\begin{equation} 
e^+ + p\ \rightarrow \
{e^+} + p' + X(j_1 + j_2 + X'), 
\end{equation} 
in which the positron is scattered at very small angles, implying 
that the emitted photon has a very low momentum ($Q^2\approx 0$) and 
can be considered, in a good approximation, as real. In such a 
context, the positron acts just as a source of photons that are 
emitted with a certain energy spectrum.  This emission can be 
expressed in terms of a ``photon flux" through the so-called 
Equivalent Photon Approximation (EPA) \cite{greiner} (or by the best 
known Weizs\"acker-Williams approximation).

There are experimental evidences that photoproduction processes take
place by two mechanisms whose calculation is considered in terms of:
(1) {\it the direct component}, in which the coupling of the photon 
to partons of the proton is point-like; and (2) {\it the resolved 
component}, in which the photon fluctuates to a partonic structure 
whose constituents interact with the partons of the proton. In fact, 
photoproduction of dijets is one of main processes by which the 
photon structure function is measured because this reaction is quite 
sensible to the quark/gluon content of the photon, even in leading 
order.

In the case of {\it diffractive} photoproduction, according to the IS
model, it is the partonic structure of the Pomeron that is probed by 
the photon itself (in the direct process) or that is envolved in 
interactions with photon constituents (in the resolved process).

At this point, it is interesting to note that the way by which the IS
model \cite {ingelman} was conceived  to describe hard diffractive
production is in complete analogy to the EPA in photoproduction
processes. Just as the electron (or positron) in photoproduction, the
proton in a diffractive interaction is scattered at very small angles 
and practically does not take part in the effective reaction. In a 
analogous way to the emission of photons and to the idea of 
(equivalent) photon flux, it sounds natural to talk about Pomeron 
emission and ``Pomeron flux factor". However, the problem with this 
analogy is that while the photon flux has a well-based theoretical 
derivation in QED, the Pomeron flux factor is obtained in a totally 
phenomenological way. This issue is central and will motivate more 
discussion below.

Back to photoproduction, the momentum distribution of the interacting
object that comes from the positron vertex (namely, the photon itself
or its constituents) is given by
\begin{eqnarray}
\label{flux_fot}
x_a f_{a/e}(x_a, \mu^2)=\int dQ^2 \int dx_\gamma \int dy\ G(y,Q^2)\ 
x_\gamma f_{a/\gamma }(x_\gamma , \mu^2)\ \delta(x_\gamma 
-{x_a}/{y}),
\end{eqnarray} 
where the photon emission is described by the flux $G(y,Q^2)$, 
obtained in the EPA context, and $x_\gamma f_{a/\gamma}(x_\gamma , 
\mu^2)$ is the photon structure function, with $x_\gamma$ being the 
fraction of the photon momentum carried by partons.

The derivation of the photon flux can be found elsewhere (see, for 
instance, \cite{greiner}) and its integrated form reads 
\begin{equation}
G(y)\ \equiv \int dQ^2\ G(y,Q^2) \ \simeq \frac{\alpha}{2\pi 
y}\left\{ [1+(1-y)^2]\ln{\frac{Q^2_{max}}{Q^2_{min}}}-2(1-y)\right\},
\end{equation}
with $Q^2_{max}$ given by the experiment and $Q^2_{min}= m_e^2\
y^2/(1-y)$, so that Eq.~(\ref{flux_fot}) can be written as
\begin{eqnarray}
\label{convof}
x_a f_{a/e}(x_a, \mu^2)= \int dy\ G(y)\ \frac{x_a}{y} 
f_{a/\gamma}(\frac{x_a}{y} , \mu^2). 
\label{fotofe}
\end{eqnarray}

\subsubsection{Cross section for the resolved component}

The cross section for diffractive photoproduction relative to the 
resolved component can be obtained in an analogous way to the 
hadroproduction expression, Eq.~(\ref{dsigjato}), but using   
\begin{equation}
x_a = y\ x_\gamma = \frac{E_T}{\sqrt{s}}(e^{\eta}+e^{\eta'}), 
\ \ \ \ 
x_b = \beta\ x_{\tt I\! P} = \frac{E_T}{\sqrt{s}}(e^{-\eta}+e^{-\eta'}), 
\end{equation}
and Eq.~(\ref{fotofe}) so that 
\begin{equation}
\label{fotoeta}
\frac{d\sigma}{d\eta}=\int dE_T^2 \int d\eta' \int dy\ G(y) \int 
dx_{\tt I\! P} \ g(x_{\tt I\! P})\ \beta f_{b/{\tt I\! P}}(\beta, 
\mu^2)\ x_\gamma f_{a/\gamma}(x_\gamma , \mu^2) \ 
\frac{d\hat{\sigma}}{d\hat{t}},
\end{equation}
with integration limits established likewise (the limits for the 
variable $y$ are given by  the experiment).

Diffractive photoproduction data are also given in terms of other 
cross sections,  ${d\sigma}/{dE_T}$, ${d\sigma}/{dW}$, 
${d\sigma}/{d\beta}$ and ${d\sigma}/{dx_\gamma}$, whose explicit 
expressions can be obtained from Eq.~(\ref{fotoeta}) with some 
appropriate change of variables.

\subsubsection{Cross section for the direct component}

The cross section corresponding to the direct component is obtained  
just by replacing the photon structure function by 
\begin{equation}
f_{a/\gamma}(x_\gamma, \mu^2)\ =\ \delta(1\ -\ x_\gamma)
\end{equation}
in Eq.~(\ref{fotoeta}) and in the other cross sections for the 
resolved component.

Note that, in this case, the Bjorken variable for the elementary 
process in leading order is $x_a=y$. Another consequence is that 
there is no direct component for ${d\sigma}/{dx_\gamma}$ in leading 
order. This will have important implications in the comparison of 
theoretical calculations with experimental data.

\section{Results and Discussion}

In the following, we present the results of our calculations of hard
diffractive production processes whose cross sections were discussed
in the previous section. For all of them, we have considered the four
possibilities of Pomeron structure function discussed in Section II. 
As for the proton and photon structure functions, we have used the 
GRV (Leading Order) parametrizations \cite{gluck,gluck1}.

\subsection{Diffractive $W$'s and the CDF rate}

As mentioned before, the diffractive $W$ production has been studied 
at the Tevatron Collider (${\sqrt s} = 1.8$ TeV) by the CDF 
Collaboration \cite{wcdf} in terms of a particular decay mode, that 
is $W \rightarrow e\ \nu$. The measurements were  triggered by 
electrons (or positrons) with transversal energy $E_T > 20$ GeV in 
the central region, $|\eta|< 1.1$, and corresponding to 
$x_{\tt I\! P} \leq 0.1$.

In Fig.~1, we show the results obtained with 
Eqs.~(\ref{dsw})-(\ref{tw}) for the STD flux factor (upper part) and 
for the REN flux factor (lower part). The calculations were performed 
with the CDF kinematical inputs, assuming that the Pomeron is emitted 
by a proton directed towards the right hand side. Due to this fact 
the $e^{\pm}$ distributions are boosted towards negative rapidity 
leaving an empty space (the characteristic rapidity gap) for $\eta > 
1.5$. In the same figure, we show the cross section for 
non-diffractive $W$ production for comparison.

It is clearly seen in Fig.~1 that the diffractive $W$ production is 
more abundant for the STD flux than it is for the renormalized one in 
spite of the fact that the Pomeron structure function applied to the 
latter case is much richer in quarks (see Section II).

In Table \ref{wrates}, we present the ratios of diffractive to
non-diffractive production rates calculated with the different
combinations of flux factor with  ${\tt I\! P}$ structure function in 
comparison with the experimental value. The theoretical ratios were 
obtained by integrating the cross sections shown in Fig.~1 over the 
CDF limits, $-1.1 < \eta < 1.1$. As can be seen, the calculations 
with the STD flux factor overestimate the experimental rate by factor 
around 3. The results obtained with the REN flux factor, on the 
contrary, are much closer to the experimental value although a little 
 below.

%
%

\subsection{Diffractive jets and Tevatron data}

Diffractive dijets rates were measured at the Tevatron Collider by  
the CDF \cite{rapgap,roman} and the D0 collaborations \cite{d01800}. 
In the CDF measurements two procedures have been used to isolate the
diffractive events, one by the rapidity gap technique \cite{rapgap} 
and the other by using roman pots to detect the recoiling proton (or 
proton remnant) \cite{roman}. The D0 analysis was performed only by 
using the rapidity gap technique, but it includes rates for two 
energies, ${\sqrt s} = 630$ GeV and  ${\sqrt s} = 1800$ GeV 
\cite{d01800}. In Table
\ref{dataprod}, the rates obtained by these experiments are shown as
well as the kinematical cuts implemented in each case. We should
notice, however, that the CDF rate given in the column (a) is the 
only experimental value already published, the others have to be 
taken as preliminary results.

In Fig.~2~(a)-(d), we show the inclusive diffractive jet cross 
section calculated with the Pomeron structure fuctions given by Fits 
1-4 (the kinematical cuts corresponding to each figure is identified 
by the letter in the top of the columns of Table II). The 
non-diffractive jet cross section calculated with the respective 
kinematical cuts is also shown.

A characteristic feature of these calculations is that the results
obtained with the STD flux with both fits are much higher than those
given by the REN flux, whose fits in turn produce rates practically
indistinguishable.

From these curves, we have the ratios of diffractive to 
non-diffractive production rates given in Table \ref{jetrates} to be 
compared with the experimental values. Again the values obtained with 
STD flux are much larger than the actual measurements while the 
results given by the REN flux are in general agreement with the 
experiments.

%
%

\subsection{Diffractive jets and ZEUS data}

The experimental data of diffractive photoproduction of jets used in
this analysis were obtained by the ZEUS Collaboration at the HERA
experiment \cite{fotopZ}, with the energy of the $\gamma^* p$ system
between the limits $134\ \leq W\ \leq 277$ GeV and with the photon
virtuality restricted by $Q^2\ \leq 4\ \rm{GeV}^2$. Other kinematical
variables that specify the outcomes of this experiment are the
following:  $\ -1.5\ \leq \eta^{lab}_{jet}\ \leq 1$, $E^{jet}_T \geq
6\ \rm{GeV}$, and $0.001\ \leq\ x_{\tt I\! P}\ \leq 0.03$. Due to the
asymmetry between the positron and proton beam energies, $E_e^{lab} 
=27.5\ \rm{GeV}$ and $E_p^{lab}=820\ \rm{GeV}$ (which correspond to
$\sqrt{s}=300\ \rm{GeV})$, the rapidity variable in the 
center-of-mass system (cms) is given by
$\eta^{cms}=\eta^{lab}+(1/2)\ln({E_e^{lab}}/{E_p^{lab}})$.

As mentioned before, the experimental differential cross
sections are given in terms of ${d\sigma}/{d{\eta}_{jet}}$,
${d\sigma}/{dW}$, ${d\sigma}/{dE_T}$, ${d\sigma}/{d\beta^{obs}}$,
and  ${d\sigma}/{dx_\gamma^{obs}}$. The superscript in the variables
$\beta^{obs}$ and $x_\gamma^{obs}$ indicates that these are 
quantities not directly measurable, but instead are {\it observables} 
obtained from the jets kinematics (see details in \cite{fotopZ}).

In Fig.~3, we show the results of ${d\sigma}/{d{\eta}_{jet}}$ for 
both STD and REN flux factors.  Here appears a situation partially 
different from what we have seen in the previous cases: the results 
obtained with the STD flux continues to overestimate the measured 
cross section by a large extent, but those obtained by the REN flux 
are not compatible with the data as it was previously observed in 
diffractive hadroproduction of $W$'s and jets.

Basically the same features are seen in Figs.~4(a)-4(d), in which we
present ${d\sigma}/{dW}$, ${d\sigma}/{dE_T}$,
${d\sigma}/{d\beta^{obs}}$, and ${d\sigma}/{dx_\gamma^{obs}}$. It is 
important to notice, however, that the curve whose shape most 
correspond to the  $\beta$-distribution in Fig.~4(c) is that one 
obtained with super-hard gluons (Fit~2). An additional observation 
about the results of Fig.~4(d) is that the cross section for this 
case does not include the direct component since our
calculation are performed only to leading order. It is known,
however, that in next-to-leading order the direct component presents 
appreciable contribution for $x_\gamma>0.75$ \cite{cimento}.

\subsection{Discussion}

Taking as whole, the results presented in Tables I and III, and in
Figs.~1-4 show that the combination STD flux factor with the Pomeron
structure functions obtained from Fits 1 and 2 cannot describe the
diffractive production of $W$'s and jets discussed here. These 
results are systematically above the data, sometimes by an order of 
magnitude or more. Thus, a natural conclusion seems to be that the 
IS model with the STD flux factor is ruled out by the experimental data.

The other general observation is that the results obtained with the 
REN flux factor and Fits 3 and 4 are in a reasonable agreement with 
the diffractive hadroproduction data, but the same theoretical scheme 
fails to describe diffractive photoproduction. 

Since the renormalization procedure is usually presented as a way
of reconciling the Ingelman-Schlein approach with the experimental
observation (and, in fact, it seems to work in hadroproduction), the
question is why such failure happens in diffractive photoproduction. 
The explanation is in the renormalization factor itself and is given 
in the following.

As pointed out in Section II, for the case of diffractive DIS, the
lower limit that enters in the definition of the (re)normalization 
term, Eq.~(\ref{norm}), is $x_{{\tt I\! P}_{min}} = Q^2/\beta s$.  
In diffractive photoproduction, we have a (dynamically) similar 
process, except for the range of values assumed by $Q^2$, implying 
that the definition of $x_{{\tt I\! P}_{min}}$ should be the same.  
In order to be clearer about the way the Pomeron structure function was 
obtained in the renormalized case, we rewrite Eq.~(\ref{Fdiff}) as
\begin{eqnarray} 
F_{2}^{D(3)}(\beta,Q^{2},x_{\tt I\! P})=g_{REN}(x_{\tt I\! P}, 
x_{{\tt I\! P}_{min}})\ F_{2}^{ {\tt I\! P}}(\beta, Q^{2}), 
\label{FRdiff} 
\end{eqnarray} 
meaning that, in such a case, the $Q^2$-dependence comes from the 
DGLAP evolution {\it and} from $x_{{\tt I\! P}_{min}}$ implicit in 
the renormalization factor.

In all of the cross section calculations presented above (including,
of course, the diffractive ones), we have applied the procedure usual
 in QCD/parton model of using $E_T$ as the evolution scale in the 
structure functions. However, in order to be consistent with the 
original proposal \cite{dino}, in diffractive photoproduction 
calculations we must assign to the
$Q^2$-dependence that belongs to renormalization factor, the 
$Q^2$-values
referring to the ZEUS experiment \cite{fotopZ}.

In fact, there is a problem about which value to choose since $Q^2 
\leq 4\ {\rm GeV}^2$, with the median value in $Q^2 \approx 10^{-3}\ 
{\rm GeV}^2$ \cite{fotopZ}.  Even in disagreement with the data, the
curves shown in Figs.~3 and 4 represent the most favorable situation,
 which corresponds to put $Q^2 = 4\ {\rm GeV}^2$. If one applies the 
median $Q^2$, these results are reduced to a really small fraction, 
five times lower than the curves presented. This is easily understood
 from the fact that  $Q^2 \sim 0$ in this kind of experiment and the 
smaller $Q^2$ is, the larger the normalization factor becomes, 
reducing considerably the calculated cross section.

\section{Hard Diffraction: Clues to a new approach}

\subsection{Diffractive Parton Model}

We propose here a new version of the Ingelman-Schlein model that,
in our view, seems to be able of overcoming the difficulties 
presented and discussed above.

First of all, we would like to state that the Pomeron flux factor, as
it is presently established, is an ill-defined and misleading 
quantity that cannot be supported only by the analogy with the 
photon flux factor (which seems to be best justification for it) or 
something alike. The concept is interesting, but its definition in 
terms of the (standard) Triple Pomeron Model leads to wrong results 
(as we have shown). Maybe, in the future, QCD will provide a rigorous
 definition for the Pomeron flux factor but, at the moment, we see no 
reason to keep it.

We think that, starting from the idea that the Pomeron is constituted of 
quarks and gluons, what one really needs to {\it estimate} the 
measured hard diffraction cross sections is a {\it probability 
distribution}  that would connect hard interactions that occur at 
partonic level to the hadronic level at which diffractive processes 
are detected. We propose that such a distribution be given by the 
{\it normalized} function
\begin{equation}
F_{sd}(x_{\tt I\! P}, t) \equiv \frac{1}{\sigma^{exp}_{sd}}\ 
\frac{d^{2}\sigma_{sd}^{exp}}{dt\ dx_{\tt I\! P}},
\label{diffac} 
\end{equation}
where ${\sigma^{exp}_{sd}}$ represents the single diffractive cross
section integrated over only one hemisphere. The other term is, of 
course, the differential cross section, that we assume to be known 
and that is, in principle, in agreement with the experimental data 
(that is what the superscript $exp$ means).

Let us call this quantity defined by Eq.~(\ref{diffac}), 
$F_{sd}(x_{\tt I\! P}, t)$, {\it diffraction factor} since it 
represents the probability distribution that a  diffractive 
interaction takes place. Once the diffraction factor is
known, we propose that the cross section of hard diffraction 
processes is the result of the convolution product
\begin{equation}
\left(\frac{d^{2}\sigma}{dx_{\tt I\! P}\ dt}\right)_{HD} = 
F_{sd}(x_{\tt I\! P}, t) \otimes {\hat{\Sigma}}_{hard},
\label{sigHD} 
\end{equation}
in which ${\hat{\Sigma}}_{hard}$ stands for all the elementary cross 
sections involved in the specific process under consideration. 
Operationally, Eq.~(\ref{sigHD}) represents what has already been 
done in Section III,  if one replaces the Pomeron flux factor by the 
diffraction factor here introduced, since the convolution product is 
conceived to be taken in the same sense of Eq.~(\ref{convol}).

Eq.~(\ref{sigHD}) is, of course, reminiscent of the IS expression, 
with exception of the normalization that, in this case, is 
established by ${\sigma^{exp}_{sd}}$ instead of the Pomeron-proton 
cross section, ${\sigma_{{\tt I\! P} p}}$, which is the original 
assumption (Cf.~\cite{ingelman}). However, this small change implies 
in two important differences: 1) $F_{sd}(x_{\tt I\! P}, t)$ is a 
normalized distribution by construction, and 2) ${\sigma^{exp}_{sd}}$ 
is an experimentally observable quantity while ${\sigma_{{\tt I\! P} 
p}}$ is a model dependent one. In order to have a brief form to 
refer to it, we are going to call the combination of 
Eq.~(\ref{sigHD}) with Eq.~(\ref{diffac}) Diffractive Parton Model 
(DIFFPM). 

Now, we intend to show that the renormalized flux factor is nothing 
but an approximate expression for the diffraction factor 
$F_{sd}(x_{\tt I\! P}, t)$. In order to do that, let us turn  our 
attention to the single diffractive cross section as it is given by 
the standard Regge theory,
\begin{equation}
\frac{d^{2}\sigma_{STD}}{dt\ dx_{\tt I\! P}} = f_{STD}(x_{\tt I\! 
P}, t) \  {\sigma}_{{\tt I\! P}p},
\label{sigSTD} 
\end{equation}
which is known from long ago to violate unitarity. Let us assume that
the corrections necessary to make this cross section compatible with
the data are (in principle) known from some physical effects 
(screening
corrections, flavoring, etc.) and that they can be represented by a 
function $C(s)$, which depends only on the energy, such that
\begin{equation}
\frac{d^{2}\sigma_{sd}^{exp}}{dt\ dx_{\tt I\! P}} = \frac{1}{C(s)}\ 
\frac{d^{2}\sigma_{STD}}{dt\ dx_{\tt I\! P}}.
\label{hyp} 
\end{equation}
Function $C(s)$ must not be confused with the renormalization factor 
$N(x_{{\tt I\! P}_{min}})$ at this point. 

Implicit in Eq.~(\ref{hyp}), there is the assumption that the 
$x_{\tt I\! P}$ and $t$ dependences given by the STD model are in 
agreement 
with the data and that the real problem has to do only with the 
energy dependence. This assumption seems to be supported by the 
analysis presented in \cite{dino2}.

For simplicity of reasoning, let us momentarily assume that the
Pomeron-proton cross section is constant, ${\sigma}_{{\tt I\! P}p} =
{\sigma}_{0}$ (which, in fact, it  approximately is). From these
hypotheses, we can extract two results:

{\bf Result 1:} By replacing Eq.~(\ref{hyp}) with (\ref{sigSTD}) into 
Eq.~(\ref{diffac}), one obtains 
\begin{equation}
F_{sd}(x_{\tt I\! P},t)=\frac{f_{STD}(x_{\tt I\! P},t)}
{\int_{x_{{\tt I\! P}_{min}}}^{x_{{\tt I\! P}_{max}}} 
\int^{0}_{t=-\infty}f_{STD}(x_{\tt I\! P},t)\ dx_{\tt I\! P}\ dt}, 
\label{diffac2}
\end{equation}
that is the same expression of the renormalized flux factor,
Eq.~(\ref{diflux}), but in which it is imposed that $x_{{\tt I\! 
P}_{min}} = (m_p + m_{\pi})^2/s$ always, that is by definition.

{\bf Result 2:} Now, by integrating Eq.~(\ref{hyp}), one gets  
\begin{equation}
\sigma_{sd}^{exp} = \frac{\sigma_0}{C(s)}\ 
{\int_{x_{{\tt I\! P}_{min}}}^{x_{{\tt I\! P}_{max}}} 
\int^{0}_{t=-\infty}f_{STD}(x_{\tt I\! P},t)\ dx_{\tt I\! P}\ dt},
\label{inthyp} 
\end{equation}
from which we see that $C(s)$ will be the same as $N(x_{{\tt I\! 
P}_{min}})$ if (and only if) $\sigma_{sd}^{exp} = \sigma_0$, and 
that means $\sigma_{sd}^{exp}$ constant. From this reasoning, one can 
obtain the renormalized expression for soft diffraction in two 
steps: firstly, one assumes that $\sigma_{sd}^{exp} = \sigma_0$ and 
determines $C(s)$ from Eq.~(\ref{inthyp}), and secondly, one replaces 
$C(s)$ so obtained in Eq.~(\ref{hyp}). In the resultant expression,
${\sigma}_{{\tt I\! P}p}$ is not  assumed to be constant anymore, 
but a constant factor, that is part of it, is adjusted according to the 
ISR data (Cf.\cite{dino}).

Let us discuss these results, starting from the second. By looking at 
the energy dependence of the $\sigma_{sd}^{exp}$ data\footnote{We 
remind the reader that the experimental data of single diffractive 
cross section are conventionally established as $\sigma_{SD}=2\ 
\sigma_{sd}^{exp}$.}, we see two different behaviors: from low 
energies up to the ISR energies, the cross section is clearly 
increasing, but from the ISR to the Tevatron energies, it is 
pratically constant (although data are really scarce in
such a region). The latter is the region in which the renormalization
scheme is applied, which is consistent with the above argumentation. 
In fact, the energy dependence obtained for the renormalized cross 
section is really mild, changing very little over a range of 
practically $10^4$ GeV (see Fig.~1 of \cite{dino}).

Of course, all of this is valid only under the supposition that
$\sigma_{sd}^{exp}$ follows this almost constant trend also in the
empty region between the ISR and the Tevatron data. If that is not 
true and the cross section have some strong variations in this 
non-observed region and/or beyond the Tevatron energy, then the 
renormalization scheme is not valid for soft diffraction anymore. In 
this case, it would be necessary a function $C(s)$ different from the
 renormalization factor $N(x_{{\tt I\! P}_{min}})$, that would 
represent such variations. But, independently on that function, the 
diffraction factor given by Eq.~(\ref{diffac2}) (in other words, the 
renormalized flux factor) would remain the same. Therefore, so far 
the conclusion is: even if the renormalization procedure were not the
 correct solution for the unitarization of the soft diffractive cross 
section, the renormalized flux factor would remain valid as an 
approximate expression for $F_{sd}(x_{\tt I\! P}, t)$.

\subsection{Application to diffractive photoproduction}

The preceding discussion allows us to change the line of 
argumentation and the way of looking at the theoretical results 
presented here so as to put in evidence the Diffractive Parton Model 
given by Eqs.~(\ref{diffac}) and (\ref{sigHD}).  From this 
point-of-view, we have already shown that the DIFFPM was able to give 
a reasonable description of the diffractive hadroproduction of $W$'s 
and jets through an approximate expression for the diffraction 
factor given by Eq.~(\ref{diffac2}), or in other words, by the
renormalized flux factor.

Now we are going to show that, despite the difficulties pointed out
previously, it is possible to give a reasonable description for
diffractive photoproduction with the same parametrizations for the
Pomeron structure function obtained with REN flux factor in
\cite{covolan}, but by applying the DIFFPM. In order to explain how that is
possible, we need to consider the renormalization procedure again.  For
the sake of simplicity, instead of using the full expression given by
Eq.~(\ref{renorm}), let us take for $N(x_{{\tt I\! P}_{min}})$ the 
approximate formulas given in Ref.~\cite{dino},
\begin{equation} 
N^{SD}(x_{{\tt I\! P}_{min}}) \approx c_1\ s^{2\epsilon}
\label{nsd}
\end{equation}
and 
\begin{equation} 
N^{HD}(x_{{\tt I\! P}_{min}}) \approx c_2\ 
\left(\frac{\beta\ s}{Q^2}\right)^{2\epsilon}, 
\label{nhd}
\end{equation}
which refer to soft and hard diffraction, respectively ($c_1$ and $c_2$ 
are just constant factors).

Based on the above equations, we see that an approximate form for 
Eq.~(\ref{FRdiff}) is 
\begin{eqnarray} 
F_{2}^{D(3)}(\beta,Q^{2},x_{\tt I\! P}) \approx  
\frac{g_{STD}(x_{\tt I\! P})}{N^{HD}(x_{{\tt I\! P}_{min}})}\ 
F_{2}^{ {\tt I\! P}}(\beta, Q^{2}), 
\label{FRapp1} 
\end{eqnarray} 
which can be rewritten as 
\begin{eqnarray} 
F_{2}^{D(3)}(\beta,Q^{2},x_{\tt I\! P}) \approx  
\frac{g_{STD}(x_{\tt I\! P})}{ 
 N^{SD}(x_{{\tt I\! P}_{min}})}\ \left[ \left( 
\frac{Q^2}{\beta}\right)^{2\epsilon} F_{2}^{ {\tt I\! P}}(\beta, 
Q^{2})\right], 
\label{FRapp2} 
\end{eqnarray}
except for constant factors.

Therefore, we see that the term above between square brackets can be
reinterpreted as an effective parametrization for the Pomeron structure
function, in which the $Q^2$-dependence comes from both the
renormalization factor and the DGLAP equations. The term on the left,
$g_{STD}/N^{SD}$, represents nothing but the diffraction factor as it
is given by Eq.~(\ref{diffac2}) (integrated over $t$, of course).

The above reasoning is a simplification to understand what has actually
been done. In summary, we have considered the $Q^2$-dependence that
comes from the renormalization factor as part of the Pomeron structure
function and, as such, it has worked as the evolution scale in the
photoproduction calculations as well. By doing so, we have {\it
effectively} established $N^{SD}(x_{{\tt I\! P}_{min}})$ as the unique
renormalization factor, which is  consistent with the DIFFPM.

By applying this procedure to the formalism described in Section III-C,
we have obtained the results for diffractive photoproduction of dijets
shown in Figs.~5 and 6.  In Fig.~5, we show the results for
${d\sigma}/{d{\eta}_{jet}}$ in comparison with the ZEUS data. As
explained in Section III, these results are obtained by summing up the
direct and the resolved components. To illustrate the importance of
taking into account both contributions, we show these components for
the dashed curve as a hachured area (direct component) and as a shaded
area (resolved component). Thus, now the theoretical results are
compatible with the data and there is no ambiguity whatsoever about how to
treat the $Q^2$-dependence.

In Figs.~6(a)-6(d), we show again the results for ${d\sigma}/{dW}$,
${d\sigma}/{dE_T}$, ${d\sigma}/{d\beta^{obs}}$, and
${d\sigma}/{dx_\gamma^{obs}}$, but now obtained with this new procedure
of calculation. From Figs.~6(a) and 6(b), we could say again that
compatibility with data is achieved (less for 6(b)), but the same does
not happen for the data of Figs.~6(c) and 6(d). In order to describe
the data exhibited in Fig.~6(c), a super-hard distribution for gluons
seems to be indispensable.  Such a distribution is very likely to
affect also the results shown in Fig.~6(d), that would tend to become
harder, providing a better description for the data. However, as
mentioned earlier, next-to-leading order calculation for the direct
component would be necessary.

Even not obtainning a perfect description of the data, we think that the
combination of these results (shown in Figs.~5 and 6) with the results
previously shown for diffractive hadroproduction composes a picture
from which it is possible to appreciate the possibilities of the DIFFPM.

\section{Concluding remarks}

We have presented in this paper an analysis on hard diffractive
processes with the aim of describing the $W$ and dijet diffractive
production rates measured at the HERA and Tevatron colliders. In order
to do that, we suggested a new approach which represents a sort of
modification in the Ingelman-Schlein model.

One (obvious) weak point in the discussion presented in Section V-A is
that we do not make any attempt to determine the function $C(s)$ without 
the renormalization scheme. Furthermore, we assumed that all
corrections could be concentrated in this factorized function which 
would depend only on the energy. Of course, this represents an
over-simplification of what really happens, but the spirit was to show
by a sort of toy-model that, even not knowing everything about the
physics of these corrections, it is possible to find some acceptable
justification for the renormalized flux factor within the DIFFPM
scheme.

The real solution might be (it certainly is) something much more
elaborate like, for instance, Tan's ``flavoring'' model \cite{tan}, 
the ``damping factor" proposed by Erhan and Schlein \cite{erhan}, or 
the ``screening corrections" of Gotsman, Levin and Maor \cite{gotsman}  
(or even something else). Whatever is the ``right" solution for the 
problem of soft diffraction unitarization, the calculational scheme 
represented by Eqs.~(\ref{diffac}) and (\ref{sigHD}) would remain 
basically the same since it does not depend on the particular model 
used to describe the single diffractive cross section (as long as such 
a model be able to provide a good description of the experimental 
data).

The basic idea underlying our proposal is that the probability of
having a diffractive (= rapidity gap) event in soft or hard
diffraction can be represented, in a good approximation, by the same
function. In our approach, it is given by the {\it diffraction factor}
here defined. We have shown that, with this idea, it is possible to 
give an acceptable description of the experimental data for both 
hadroproduction and photoproduction in hard diffractive interactions.


\section*{Acknowledgments}

We would like to thank the Brazilian governmental agencies CNPq and 
FAPESP for their financial support.


\appendix

\section{Diffractive Hadroproduction of $W^{\pm}$}

\subsection{Non-diffractive W production} 

In order to calculate the cross section for the reaction 
(\ref{reacao}), we start by dealing with the expression  for the {\it 
non-diffractive} production of $W$'s, 
\begin{eqnarray}
\label{mp}
d\sigma = \sum_{a,b} \int dx_a \int dx_b\ f_{a/p}(x_a)\ 
f_{b/\bar{p}}(x_b) \ \frac{d\hat{\sigma}}{d\hat{t}}(ab \rightarrow W 
\rightarrow e\nu)\ d\hat{t}.
\end{eqnarray}
In this equation, $f_{i/A}(x_i)$ refers to the distribution of partons 
$i$ in hadron A.  In the hadron center-of-mass system (cms), the 
total, longitudinal and transversal energy of the electron
(or positron) are given, respectively, by
\begin{eqnarray}
\label{ee}
E_e &=& \frac{\sqrt{s}}{4}\ [x_a(1+\cos\theta) + x_b(1-\cos\theta)], \\
E_{L} &=& \frac{\sqrt{s}}{4}\ [x_a(1+\cos\theta) - x_b(1-\cos\theta)], \\
E_T &=& \frac{M_W}{2} \sin \theta,
\label{et}  
\end{eqnarray}
in which the constraint $\hat{s}=x_a x_b s=M_W^2$ has been used and 
where $\theta$ is the electron scattering angle with respect to the 
proton beam direction (which, in this paper, is assumed to be the 
positive $z$-direction). 
From the expression for the electron rapidity, 
\begin{eqnarray}
\label{rap}
{\eta_e} &=& \ln\left(\frac{E_e+E_{L}}{E_T}\right),
\end{eqnarray}
and using 
Eqs.~(\ref{ee})-(\ref{et}), one gets 
\begin{eqnarray}
\label{xaww}
x_a &=& e^{\eta_e}\ \frac{M_W}{\sqrt{s}}\  \sqrt{\frac{1-\cos\theta}
{1+\cos\theta}} \end{eqnarray}
and
\begin{eqnarray}
\label{xbww}
x_b = \frac{M_W^2}{x_a s} &=& e^{-\eta_e}\ \frac{M_W}{\sqrt{s}}\  
\sqrt{\frac{1+\cos\theta}{1-\cos\theta}}.
\end{eqnarray}

The Mandelstam variables for the elementary process, $a\ b \rightarrow W 
\rightarrow e\ \nu$, are
\begin{eqnarray}
\hat{s}&=&(p_a+p_b) = M_W^2, \\
\label{tchap}
\hat{t}&=&(p_e-p_a)^2 = -\frac{M_W^2}{2}\ (1-\cos\theta),  \\
\hat{u}&=&(p_e-p_b)^2 = -\frac{M_W^2}{2}\ (1+\cos\theta), 
\label{uchap}
\end{eqnarray}
and, therefore, we have $dx_b\ d\hat{t} = {x_b}/{\sqrt{A^2-1}}\ \ dE_T\ 
d\eta_e$, where $A$ is defined as 
\begin{eqnarray}
\label{A}
A \equiv \frac{M_W}{2 E_T}.
\end{eqnarray}
Such a change of variables allows one to rewrite Eq.~(\ref{mp}) as 
\begin{eqnarray}
\frac{d\sigma}{d\eta_e} &=& \sum_{a,b} \int dE_T \int dx_a\ f_{a/p}(x_a) 
\ f_{b/\bar{p}}(x_b)\ \frac{x_b}{\sqrt{A^2-1}}\ 
\frac{d\hat{\sigma}}{d\hat{t}}
(ab \rightarrow W \rightarrow e\nu).
\end{eqnarray}
Now, from Eqs.~(\ref{et}) and Eq.~(\ref{A}), one obtains
\begin{eqnarray}
\label{coseno}
\cos\theta = \pm \frac{\sqrt{A^2-1}}{A}.
\end{eqnarray}
Of course, the positive or negative signs indicate the direction in 
which the electron (or positron) is being emitted. This sign is chosen 
according to the following criterion: due to helicity conservation, 
the electron is preferentially  emitted in the proton beam direction, 
such that in the case of   $W^- \rightarrow e^-\ {\bar\nu}_e$ one 
should use
\begin{eqnarray}
\label{coseno+}
\cos\theta = + \frac{\sqrt{A^2-1}}{A}.
\end{eqnarray}

By applying such a criterion to the case of $W^-$ production, from 
Eqs. (\ref{xaww})-(\ref{uchap}) one gets   
\begin{eqnarray}
\label{xa-}
x_a = e^{\eta_e}\ \frac{M_W}{\sqrt{s}}\ \left(A - \sqrt{A^2-1}\right), \\
\label{xb}
x_b = e^{-\eta_e}\ \frac{M_W}{\sqrt{s}}\ \left(A + \sqrt{A^2-1}\right), \\
\hat{u}=-\frac{\hat{s}}{2}(1+\cos\theta)\ =\ -E_T\ M_W 
\left(A+\sqrt{A^2-1}\right), \\
\label{tc1}
\hat{t}=-\frac{\hat{s}}{2}(1-\cos\theta)\ =\ -E_T\ 
M_W \left(A-\sqrt{A^2-1}\right).
\end{eqnarray}

Similarly, in the reaction $W^+ \rightarrow e^+\ {\nu}_e$ the positron is
 preferentially produced in the antiproton direction, such that 
\begin{eqnarray}
\label{coseno-}
\cos\theta = - \frac{\sqrt{A^2-1}}{A}.
\end{eqnarray}
Thus, in the case of $W^+$ production, Eqs.~(\ref{xaww})-(\ref{uchap}) 
give
\begin{eqnarray}
\label{xa+}
x_a  = e^{\eta_e}\ \frac{M_W}{\sqrt{s}}\ \left(A+\sqrt{A^2-1}\right), \\
x_b = e^{-\eta_e}\ \frac{M_W}{\sqrt{s}}\ \left(A-\sqrt{A^2-1}\right), \\
\hat{u}\ =\ -E_T\ M_W \left(A-\sqrt{A^2-1}\right), \\
\hat{t}\ =\ -E_T\ M_W \left(A+\sqrt{A^2-1}\right).
\label{tc2}
\end{eqnarray}

The elementary cross section for $W$ production is given by
\begin{equation}
\label{sigcha}
\frac{d\hat{\sigma}}{d\hat{v}}=\frac{1}{x_b s}\ 
\delta(x_a-{M_W^2}/{x_b s})\ \frac{G_F^2}{6 M_W \Gamma_W}\ 
V_{ab}^2\ \hat{v}^2,
\end{equation}
where $M_W$ is the $W$ mass, $V_{ab}$ is the Kobayashi-Maskawa matrix 
element, $G_F$ is the Fermi  constant, and $\Gamma_W$ is the $W$ 
decay width. In Eq.~(\ref{sigcha}), variable $\hat{v}$ stands for 
$\hat{u}$ 
or $\hat{t}$ according to $W^-$ or $W^+$ production, respectively.

In summary,  the cross section for $W^-$ production is
\begin{eqnarray}
\frac{d\sigma}{d\eta_e}= \sum_{a,b} \int dE_T\ f_{a/p}(x_a)\ 
f_{b/{\bar{p}}}(x_b)\  \left[\frac{ V_{ab}^2\ G_F^2}{6 s \Gamma_W 
M_W}\right] \frac{\hat{u}^2}{\sqrt{A^2-1}}
\end{eqnarray}
for which Eqs.~(\ref{xa-})-(\ref{tc1}) apply. 
For the case of $W^+$ production, the cross section is  
\begin{eqnarray}
\label{dsig2}
\frac{d\sigma}{d\eta_e}= \sum_{a,b} \int dE_T\ f_{a/p}(x_a)
\ f_{b/{\bar{p}}}(x_b)\  \left[\frac{V_{ab}^2\ G_F^2}{6 s \Gamma_W 
M_W}\right]\ \frac{\hat{t}^2 }{\sqrt{A^2-1}},
\end{eqnarray}
with Eqs.~(\ref{xa+})-(\ref{tc2}).

\subsection{Diffractive W production} 

From the discussion above, the cross section for the diffractive case 
is easily obtained. Introducing the prescription established by Eqs. 
(\ref{convol}) and  (\ref{convoP}) into the expressions derived above, 
the cross section for diffractive $W$ production becomes
\begin{eqnarray}
\frac{d\sigma}{d\eta_e}= \sum_{a,b} \int dx_{\tt I\! P}\ g(x_{\tt I\! 
P})\int dE_T\  f_{a/{\tt I\! P}}(x_a)\ f_{b/\bar{p}}(x_b)\ 
\left[\frac{ V_{ab}^2\ G_F^2}{6 s \Gamma_W M_W}\right] 
\frac{\hat{t}^2\ (\hat{u}^2)}{\sqrt{A^2-1}}
\end{eqnarray}
\\
where $g(x_{\tt I\! P})$ is the integrated flux factor and  
\begin{eqnarray}
x_a  &=& \frac{M_W\ e^{\eta_e}}{(\sqrt{s}\ x_{\tt I\! P})}\left(A \pm 
\sqrt{A^2-1}\right), \\
x_b &=& \frac{M_W\ e^{-\eta_e}}{\sqrt{s}}\left(A \mp \sqrt{A^2-1}\right). 
\end{eqnarray}
We remind that, in the definition of these variables, it is assumed 
that the Pomeron is emitted by the proton and that $W$ production is
the result of ${\tt I\! P}-{\bar p}$ interaction (as in the CDF experiment). 
The choice of the signs and of the variables $\hat{t}$ 
and $\hat{u}$ proceeds in the same way as in the non-diffractive case.

%
%

\begin{figure}[htbp]
\centerline{\psfig{figure=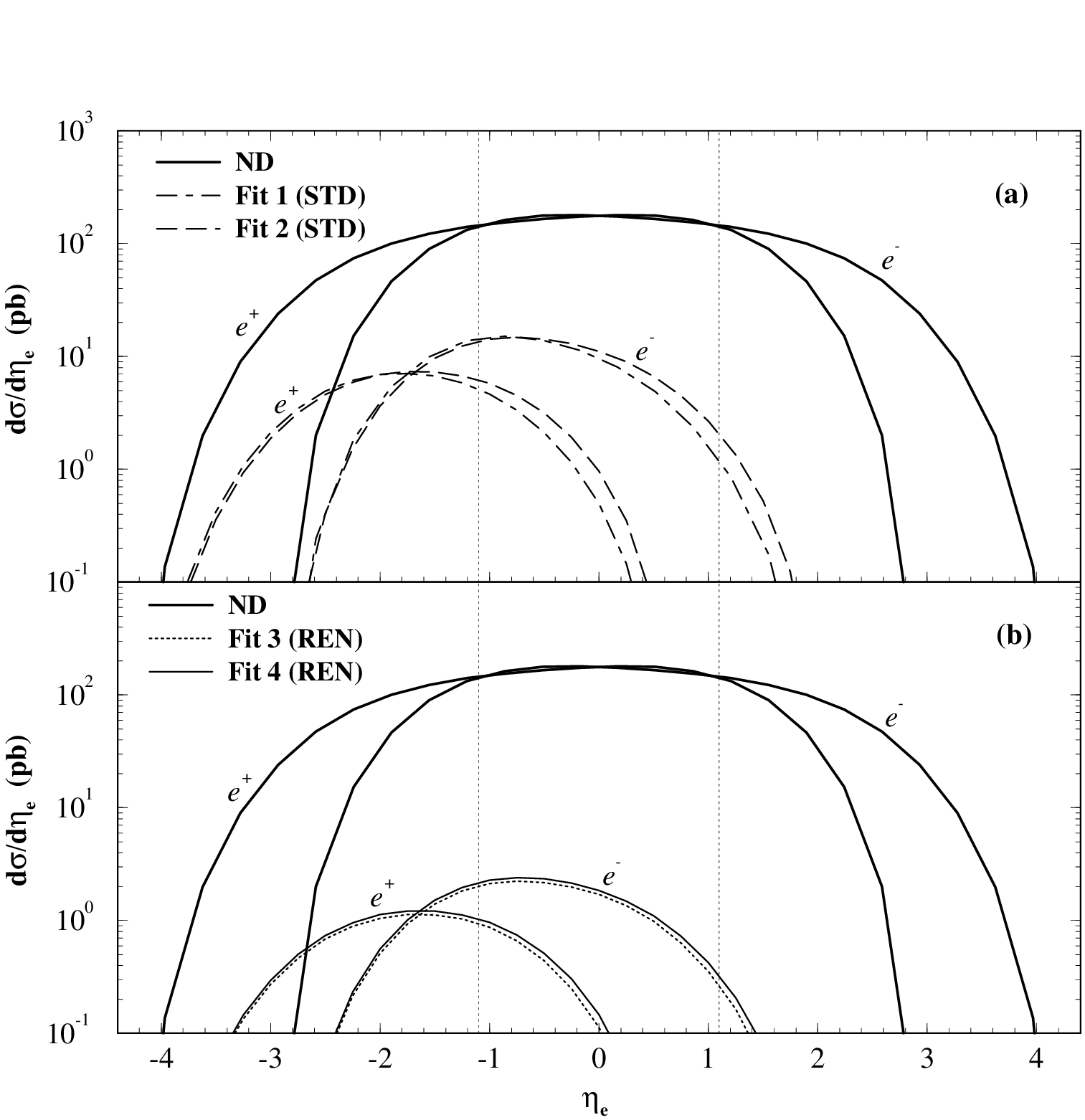,width=12cm}}
\vspace{1.6cm}
\caption{Rapidity distributions of \protect{$e^{\pm}$} emitted in
\protect{$W^{\pm}$} hadroproduction processes. The curves labeled ND refer to 
the non-diffractive cross section. The other curves correspond to the
result of diffractive production, with the labels indicating the
respective Pomeron structure function and flux factor used in the
calculation (see text). The vertical dotted lines establish the
rapidity limits within which the CDF measurements were performed.}
\end{figure}

\begin{figure}[htbp]
\centerline{\psfig{figure=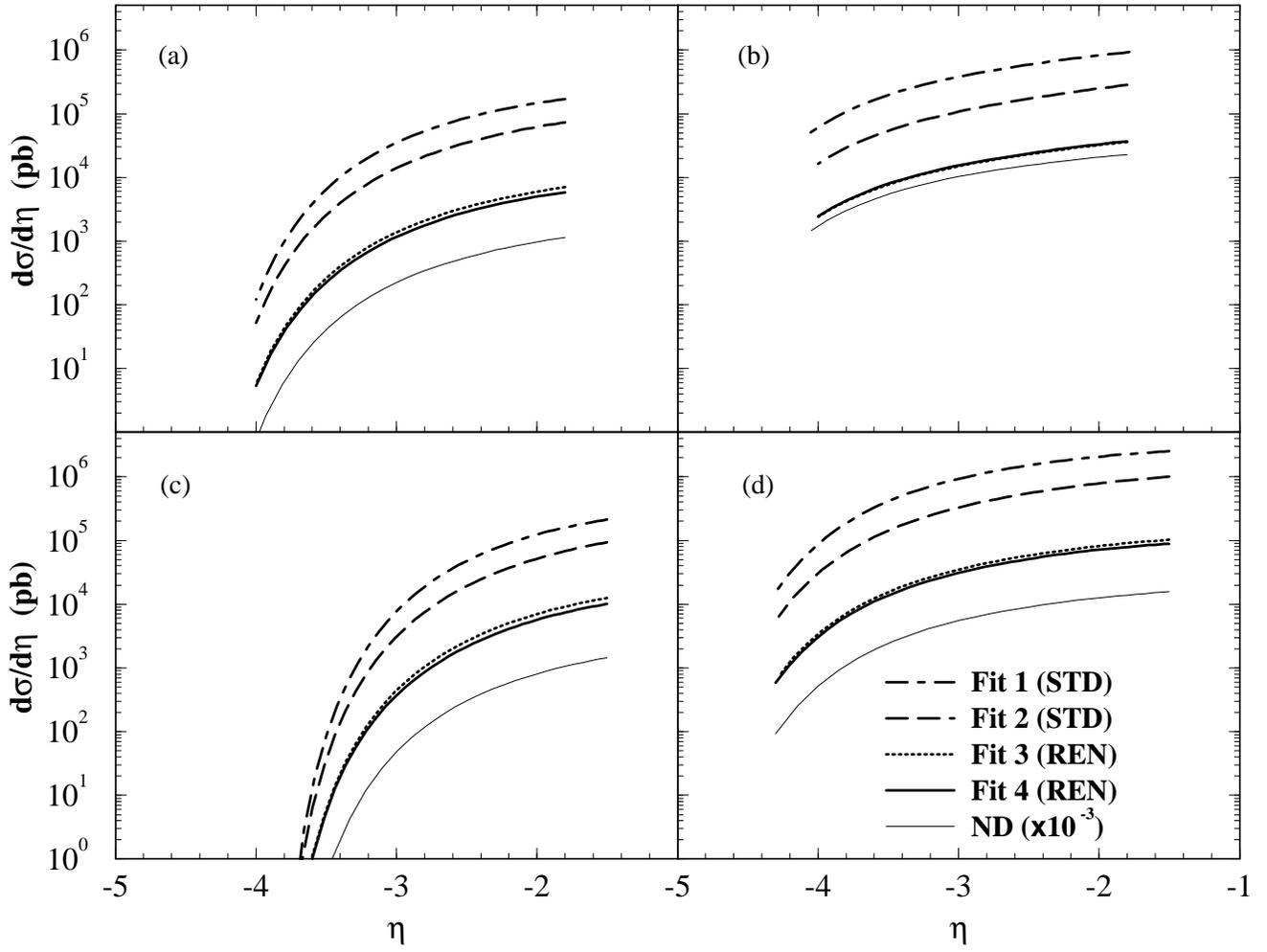,width=15cm}}
\vspace{1.6cm}
\caption{Rapidity distributions of jets from 
hadroproduction processes.  The curves labeled ND refer to the 
non-diffractive cross section (multiplied by a factor \protect{$10^{-3}$}). The 
other curves correspond to the result of diffractive production, with 
the labels indicating the respective Pomeron structure function and 
flux factor used in the calculation. The letter in the top of each 
figure indicates the corresponding kinematical cuts presented in Table
 II and applied to the calculations.}
\end{figure}

\begin{figure}[htbp]
\centerline{\psfig{figure=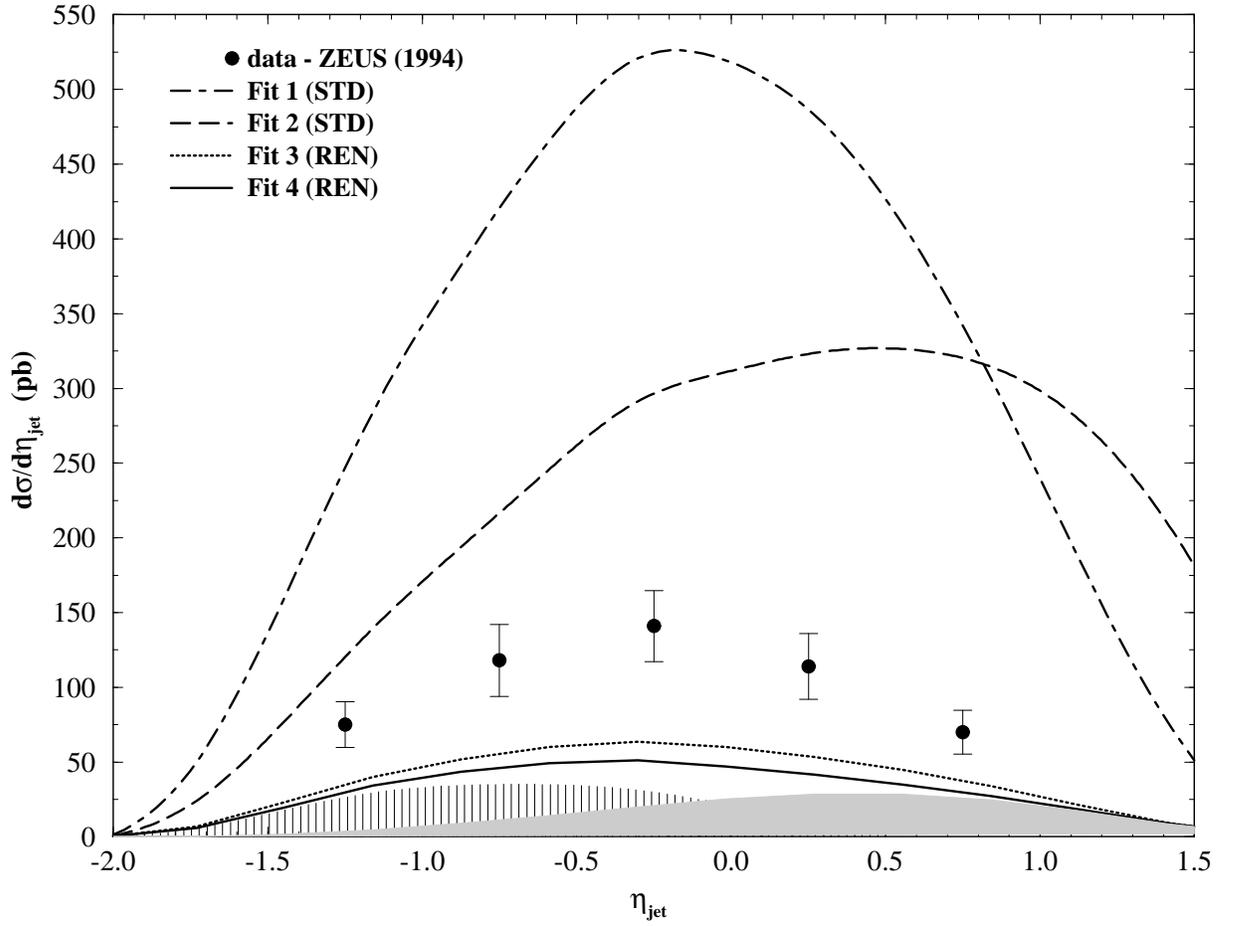,width=15cm}}
\vspace{1.6cm}
\caption{Rapidity distributions of jets from diffractive
photoproduction processes. The labels indicate
the respective Pomeron structure function and flux factor used in the
calculation. The experimental data were measured by the ZEUS
Collaboration \protect{\cite{fotopZ}}.} 
\end{figure}

\begin{figure}[htbp]
\centerline{\psfig{figure=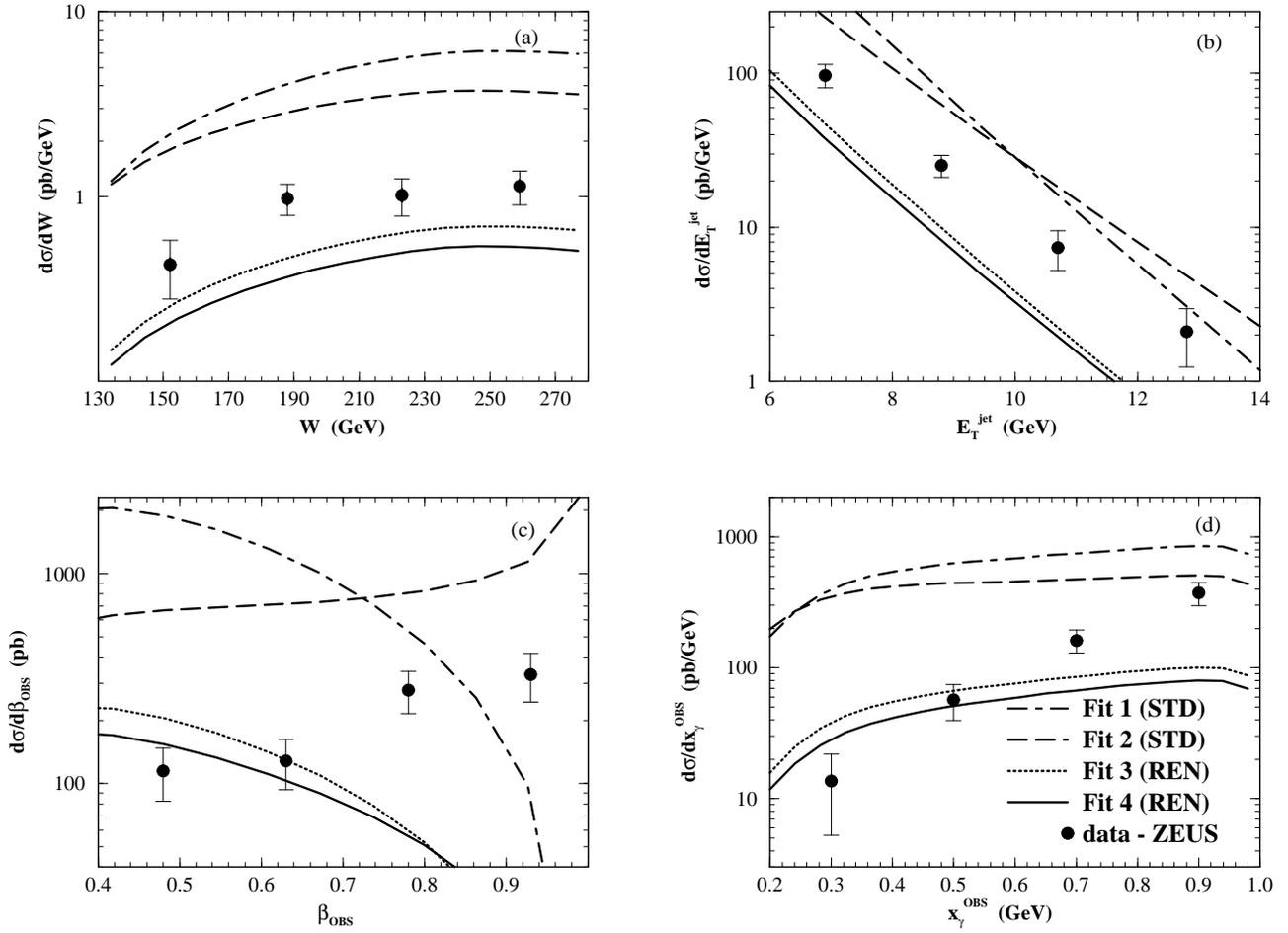,width=15cm}}
\vspace{1.6cm}
\caption{Cross sections relative to diffractive
photoproduction of jets in comparison to ZEUS data \protect{\cite{fotopZ}}. 
As in the other figures, the curve labels indicate
the respective Pomeron structure function and flux factor used in the
calculation.}
\end{figure}

\begin{figure}[htbp]
\centerline{\psfig{figure=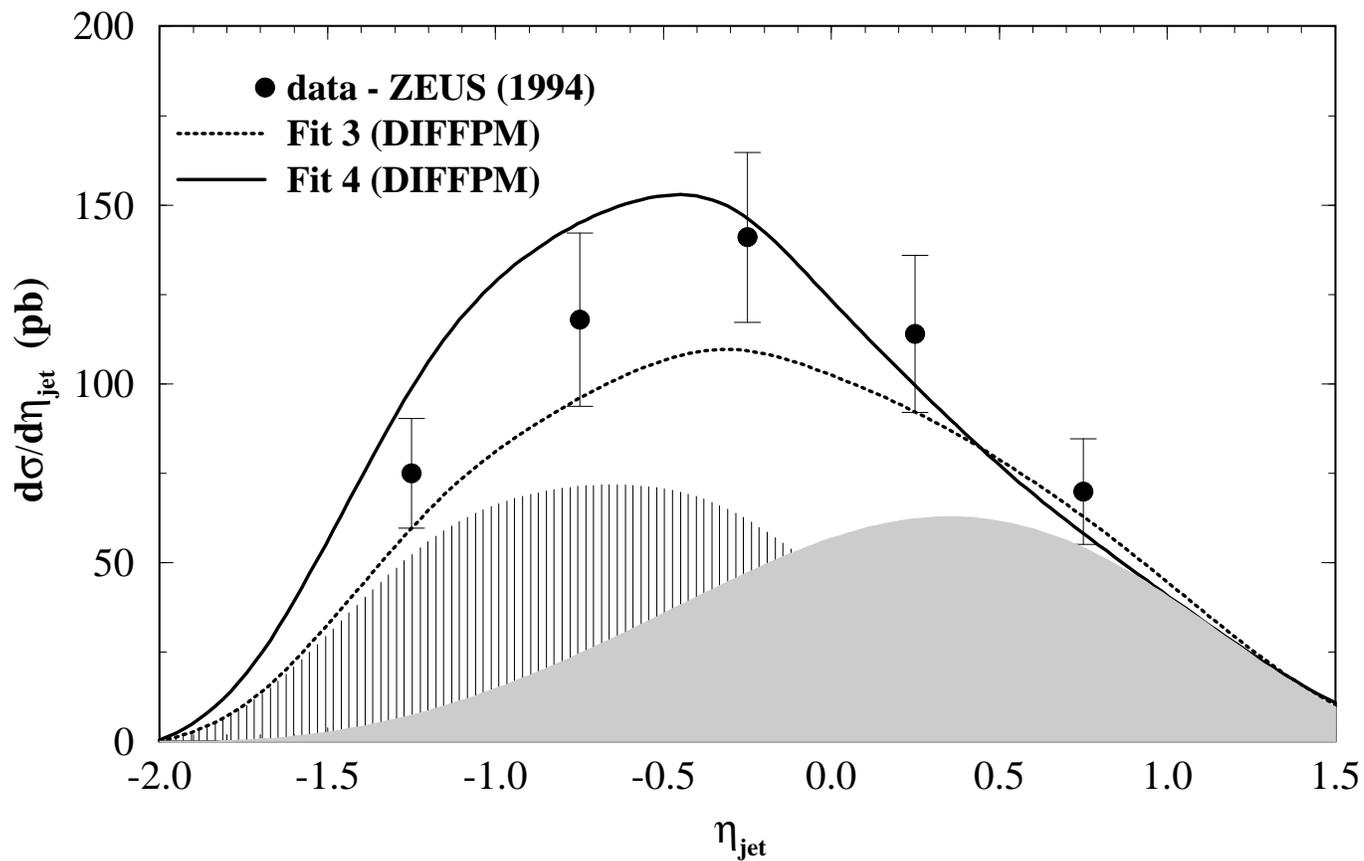,width=15cm}}
\vspace{1.6cm}
\caption{Rapidity distributions of jets from diffractive 
photoproduction processes. The labels indicate the respective Pomeron 
structure function used in the calculation, but with the redefinition 
discussed in the text. The experimental data were measured by the ZEUS
Collaboration \protect{\cite{fotopZ}}. For the dotted curve, it is also shown 
its components: the direct contribution (hachured area) and resolved
contribution (shaded area).}
\end{figure}

\begin{figure}[htbp]
\centerline{\psfig{figure=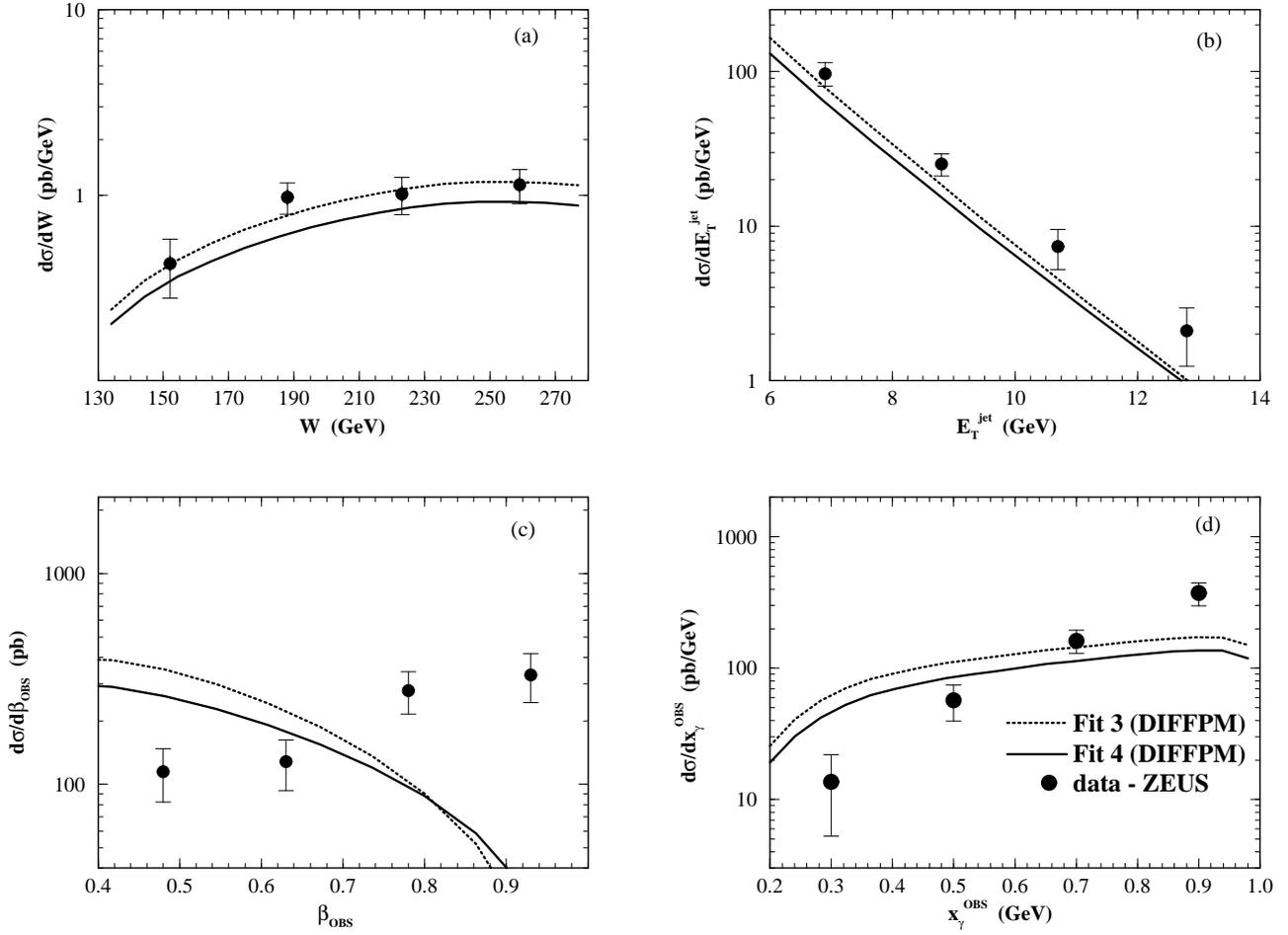,width=15cm}}
\vspace{1.6cm}
\caption{Cross sections relative to diffractive
photoproduction of jets in comparison to ZEUS data \protect{\cite{fotopZ}}.  The
Pomeron structure function used in the calculations are the same as in
Fig.~5.}
\end{figure}

\newpage

\begin{table}
\begin{tabular}{|c|c||c|c|c|c|} 
& & STD  &  STD & REN & REN \\
EXPERIMENT & RATE & Fit 1 & Fit 2 & Fit 3 & Fit 4 \\
($W$'s) & & (hard-hard) & (free-delta) & (hard-hard) & (free-zero) \\
\hline
 & & & & & \\
CDF (Rap-Gap) & $1.15\pm 0.55$ & 3.12 & 3.54 & 0.53 & 0.58   \\ 
  & & & & & \\
\end{tabular}
\vspace{0.2cm}
\caption{\sf{Diffractive production rates of $W$'s (all values are 
given in percentages).}}
\label{wrates}
\end{table}

\begin{table}
\begin{tabular}{|c|c|c|c|c|} 
 &  (a) & (b) & (c) & (d) \\ 
EXPS. &  CDF  & CDF  & D0  & D0  \\ 
 & (Rap-Gap) & (Roman Pots) & (630 GeV) & (1800 GeV) \\ \hline
 & & & & \\
RATES & $0.75\pm 0.10$  & $0.109\pm 0.016$ & 1-2 & $0.67\pm 0.05$ \\
(\%) &  & & & \\ \hline
 & & & & \\
rapidity & $-3.5 < \eta < -1.8$ & $-3.5 < \eta < -1.8$  & 
 $-4.1 < \eta < -1.6$  & $-4.1 < \eta < -1.6$ \\ 
 & & & & \\ \hline 
 & & & & \\
$x_{\tt I\! P}$ &  $x_{\tt I\! P} < 0.1 $ & $ 0.05 < x_{\tt I\! P} < 
0.1$ & $ x_{\tt I\! P} < 0.1 $ & $  x_{\tt I\! P} < 0.1 $ \\
 & & & & \\ \hline
 & & & & \\
$E_{T_{min}}$ & $20$ GeV & $10$ GeV &  $12$ GeV  & $12$ GeV \\ 
 & & & & \\ 
\end{tabular}
\vspace{0.2cm}
\caption{\sf{Experimental data and respective kinematical cuts for 
different measurements of diffractive production of dijets.}}
\label{dataprod}
\end{table}

\begin{table}
\begin{tabular}{|c|c||c|c|c|c|} 
& & STD &  STD & REN & REN \\
EXPERIMENTS& RATES & Fit 1 & Fit 2 & Fit 3 & Fit 4 \\
(JETS)& & (hard-hard) & (free-delta) & (hard-hard) & (free-zero) \\
\hline
& & & & & \\ 
CDF (Rap-Gap) & $0.75\pm 0.10$  & 15.3  & 6.33 & 0.62 & 0.52  \\  
 &  & & & & \\ \hline
& & & & & \\ 
CDF (Roman Pots)  & $0.109\pm 0.016$ & 3.85 & 1.13 & 0.15  &  0.16 \\
  & & & & & \\ \hline  
& & & & & \\ 
D0 (630 GeV) & 1-2 & 15.4 & 6.41 & 0.87 & 0.71  \\ 
 & & & & & \\ \hline  
& & & & & \\ 
D0 (1800 GeV)  & $0.67\pm 0.05$  & 16.6 & 6.14 & 0.65 & 0.57  \\ 
 & & & & & \\ 
\end{tabular}
\vspace{0.2cm}
\caption{\sf{Diffractive production rates of dijets (all values are 
given in percentages).}}
\label{jetrates}
\end{table}

\end{document}